# Magnetic Field and Displacement sensor based on Giant Magneto-impedance effect

B. Kaviraj[1] and S. K. Ghatak

*Department of Physics and Meteorology, Indian Institute of Technology, Kharagpur, India*

## Abstract

A two-core transducer assembly using a $Fe_{73.5}Nb_3Cu_1Si_{13.5}B_9$ ribbon to detect a change of magnetic field is proposed and tested for displacement (linear and angular) and current sensor. Two identical inductors, with the ribbon as core, are a part of two series resonance network, and are in high impedance state when excited by a small a.c field of 1MHz in absence of d.c biasing field ($H_{dc}$). When the magnetic state of one inductor is altered by biasing field, produced by a bar magnet or current carrying coil, an ac signal proportional to $H_{dc}$ is generated by transducer. The results for the sensitivity and linearity with displacement (linear and angular) of a magnet and with field from the current carrying coil are presented for two particular configurations of the transducer. High sensitivities of voltage response as much as 12μV/μm and 3mV/degree have been obtained for the transducer as a linear and angular displacement sensor respectively in the transverse configuration of exciting a.c and biasing d.c fields.

---
[1] Electronic address: bhaskar@phy.iitkgp.ernet.in

# 1. Introduction

The amorphous magnetic materials, due to their superior magnetic properties, are being considered as new materials for high performance sensors [1] and other devices [2]. Using amorphous ribbon, different types of sensors, for e.g. force [3-5] and displacement [6] sensors have been realized. The transducers uses either open flux configuration with two-core transformer or closed flux configuration with toroidal core. Their detection range, linearity and accuracy depend on the magnetic properties of the ribbon and geometry of the core. In a two-core multivibrator circuit, a dc output is obtained when the two cores are non-uniformly polarized. In an ac transducer [4,5,8], the core is magnetized by an ac field and an ac output signal changes when the toroidal core is subjected to an external force or magnetic field. The use of ac magnetization has the advantage over dc transducer systems [8]. In this paper we propose an ac two-core transducer, which can be used to detect small displacement (linear or angular), and current or magnetic field. The two cores constitute two identical arms of the primary of a transformer so that no signal appears at the secondary when the primary is excited by an a.c source. The signal at the secondary appears due to the impedance imbalance in the two arms, which is caused by an external magnetic field. The transition metal – metalloid based glasses exhibit Giant Magneto-Impedance (GMI) effect due to the soft ferromagnetic behaviour [9]. The impedance of these materials to inflow of small a.c. current at r.f region is very large due to large permeability in absence of any other magnetic field. The high impedance state is related to strong screening effect of electric field caused by a.c. magnetization current. In presence of biasing d.c. field, much larger than a.c. exciting field, the screening is drastically reduced resulting low impedance. This negative magneto-impedance effect can therefore be used to detect and measure magnetic field [8-14]. We report here a transducer assembly that utilizes the GMI effect in a $Fe_{73.5}Nb_3Cu_1Si_{13.5}B_9$ ribbon to detect small change in position (linear or angular) of the source of the magnetic field. The magnitude and range of linearity of the transducer signal depends on, apart from biasing field, a number of parameters like amplitude of excitation, core materials, different transducer configuration, etc. The dependence of the signal on above parameters is presented, and the possibility of using the transducer as displacement (linear or angular) and current sensor is explored.

## 2. Transducer System

The transducer assembly is shown in Fig. 1. $L_1$ and $L_2$ are two identical coils (100 turns) and are wrapped over an amorphous ribbon, which acts as core. The length of the ribbon (4cm) is larger than the length of the coils (1cm). The coils $L_1$ and $L_2$ together with two small coils $P_1$ and $P_2$ constitute two inductors as identical parallel arms between the ac source and the ground. The capacitors $C_1$ and $C_2$ are adjusted so that the two arms are at resonance at the excitation frequency.

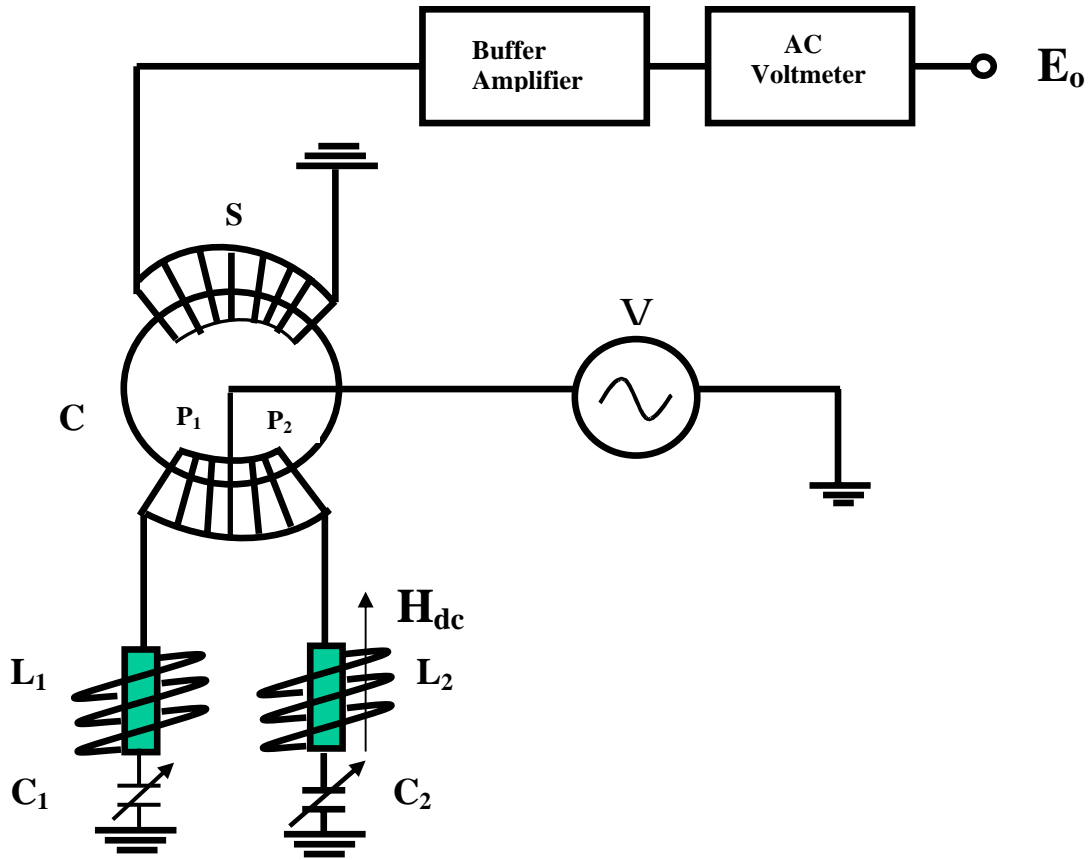

**Fig. 1**

Fig 1. Block diagram of transducer assembly. $L_1$ and $L_2$ are two identical coils with $Fe_{73.5}Nb_3Cu_1Si_{13.5}B_9$ ribbon as core. $P_1$ and $P_2$ are two identical primaries, S is the secondary and C is the ferrite core. The inductances of the primary and secondary were 1.45µH and 10µH respectively. The number of turns in $P_1$ and $P_2$ was 4 and that of secondary was 100. The values of capacitances $C_1$ and $C_2$ were 2.5nF each.

The small ac current in L's is used to magnetize the ribbon and the frequency of exciting field is kept at 1MHz for all the results. When the impedance of the two arms are identical, the net current to magnetize the ferrite core (C) of the transformer is nearly zero, and an almost zero signal appears across the secondary S. When the symmetry of the arms is disturbed by placing one inductor (say $L_2$) in an external magnetic field $H_{dc}$, the impedance of the coil is reduced due to negative magneto-impedance effect in the amorphous ribbon. This material due to magnetic softness exhibits Giant Magneto-impedance (nearly 90% change of impedance in presence of $H_{dc} \approx 100$ Oe [15] ), which is much large when compared with the Co-rich materials, which exhibits maximum GMI ratios about 40% [16]. The auxiliary ferrite core C is magnetized due to impedance imbalance of $L_1$ and $L_2$ and a signal appears at S.

The amplitude of the output signal is given by,

$$E_0 = K\, N_s\, \mu_c\, \omega\, V\, [Z_1 - Z_2]/\, Z_1\, Z_2 \quad \ldots\ldots\ldots\ldots\ldots\ldots (1)$$

where $Z_1$ ($Z_2$) is impedance of arm 1(2), $N_s$ = number of turns of secondary S and $\mu_c$ = permeability of auxiliary core and constant K depends on area of coil S, amplifier gain, number of turns of P's.

In deriving the above equation (1) it is assumed that (a) the current in the secondary is zero, (b) impedance of $P_1$ (or $P_2$) is negligibly small compared to that of $L_1$ (or $L_2$) and (c) the impedances of coils $L_1$ and $L_2$ are nearly inductive at the operating frequency. The first condition is realized by connecting the secondary S to high input impedance buffer amplifier. The number of turns $N_p = 4$, is small compared to that of 100 of $L_1$ and this assures the second condition. The frequency of the ac excitation of the transducer is kept at 1MHz throughout the measurement. This frequency is chosen to satisfy the third assumption and is not too high to induce any oscillation tendency of the amplifier. The signal $E_0$ is proportional to the inductive difference ($Z_1$-$Z_2$) and is enhanced due to the mutual coupling of the primary and secondary coil. The core C acts as a parametric amplifier. The output signal $E_0$ is measured with the help of an ac voltmeter. The external magnetic field has been applied in the longitudinal and transverse direction of the ribbon and is produced either by a permanent magnet or an energized air core coil of sensitivity 1Oe/A.

## 3. Results

In Fig (2), we depict the possibility of the transducer for detecting magnetic fields ($\leq$ 1Oe). Here the signal in the transducer is generated by a current carrying air-core coil. The coil was placed with its axis parallel to the length of the ribbon. As shown in Fig (2), the output voltage '$E_0$' is plotted as a function of external dc magnetic field. Further the voltage responses were also observed at different values of excitation currents at $i_c$ = 0.4, 1.6 and 4mA. The excitation frequency was kept constant at a value of 1MHz. The voltage responses are higher for higher values of excitation currents. Note also the dramatic increase of sensitivity of the voltage response (linear portion) with the increase of excitation current. At $i_c$ = 0.4mA, the sensitivity is about 90mV/Oe which increases to 140mV/Oe at $i_c$ = 1.6mAand finally to 260mV/Oe at $i_c$ = 4mA.

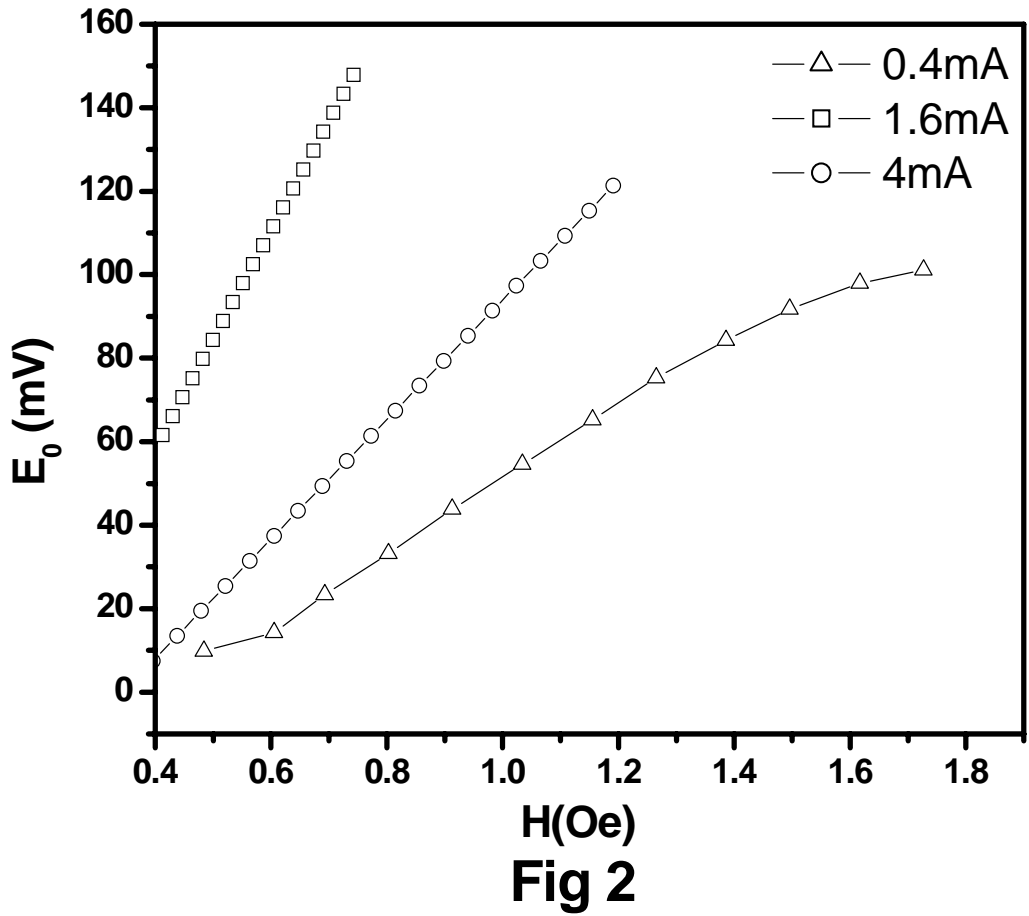

Fig 2. The output voltage 'E$_0$' vs the dc magnetic field from a Helmholtz coil in a longitudinal configuration and at excitation current amplitudes of i$_c$=0.4,1.6 and 4mA. The corresponding sensitivities of the linear portions of voltage responses to the dc field current were 90mV/Oe, 140mV/Oe and 260mV/Oe respectively.

In Fig. 3, the amplified signal output E$_0$ is plotted as a function of displacement (d) of a small cylindrical magnet (H=150Oe at one of its face and decreases sharply along its axis) and for two different values of excitation currents, i$_c$=0.4 and 1.6mA.

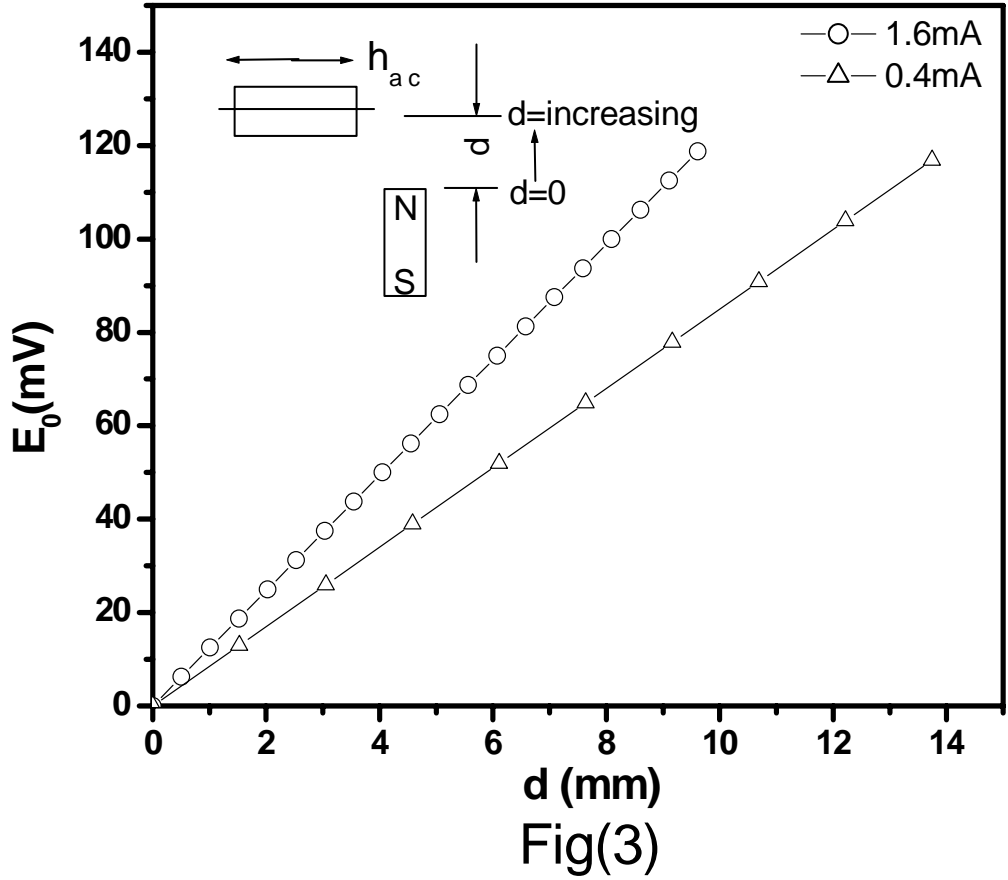

Fig(3)

Fig 3. Plot of output voltage 'E$_0$' vs the displacement 'd' of a small magnet with respect to a core L$_2$, as shown in inset, for excitation currents, i$_c$= 0.4mA and 1.6mA. In this configuration the exciting ac field is nearly transverse to biasing dc field.

In this configuration, the biasing d.c field acting on one of the cores (as shown in the inset) is  nearly transverse to the exciting a.c field. The presence of H$_{dc}$ reduces the response of the ribbon to the a.c field. The reduction of the permeability µ, in turn, increases the screening length δ (proportional to µ$^{-1/2}$ ) of electromagnetic  field. This leads to impedance imbalance between the coils. As the excitation voltage is kept fixed, the differential current in two arms produces signal at secondary S. A linear increase of $E_0$ with d is obtained as the magnet is brought closer to one end of the ribbon. Also the values of $E_0$ are higher for larger values of excitation currents. For $i_c$ = 0.4mA, the sensitivity of the voltage response is 8.5µV/µm which increases to almost 12.4 µV/µm as the magnitude of excitation current is increased.

In Fig 4, the results for the voltage response where exciting and biasing fields are collinear (longitudinal configuration) are shown as a function of displacement.

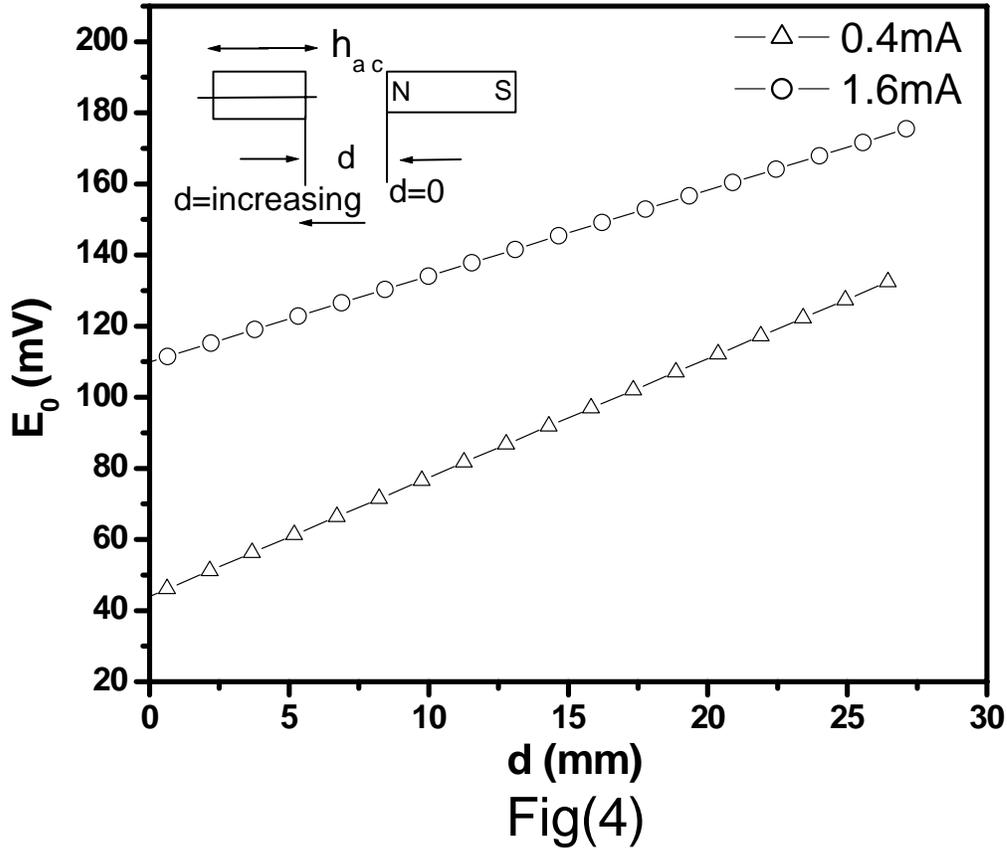

Fig 4. Plot of output voltage '$E_0$' vs the displacement 'd' of a magnet in longitudinal configuration (h parallel to $H_{dc}$) at excitation currents, $i_c$= 0.4mA and $i_c$=1.6mA.

Here also the output voltage increases linearly with 'd' and the magnitudes of the voltage response are higher for higher values of excitation currents. In this configuration, it has been observed that the sensitivity at lower value of excitation current ($i_c$ = 0.4mA) is 3.3 µV/µm, which is slightly higher than the sensitivity at higher value of excitation current ($i_c$ = 1.6mA), which is 2.4 µV/µm.

Fig 5 depicts the angular variation of $E_0$ when the bar magnet placed at a distance d=5mm from the ribbon end is given an angular displacement from its initial position.

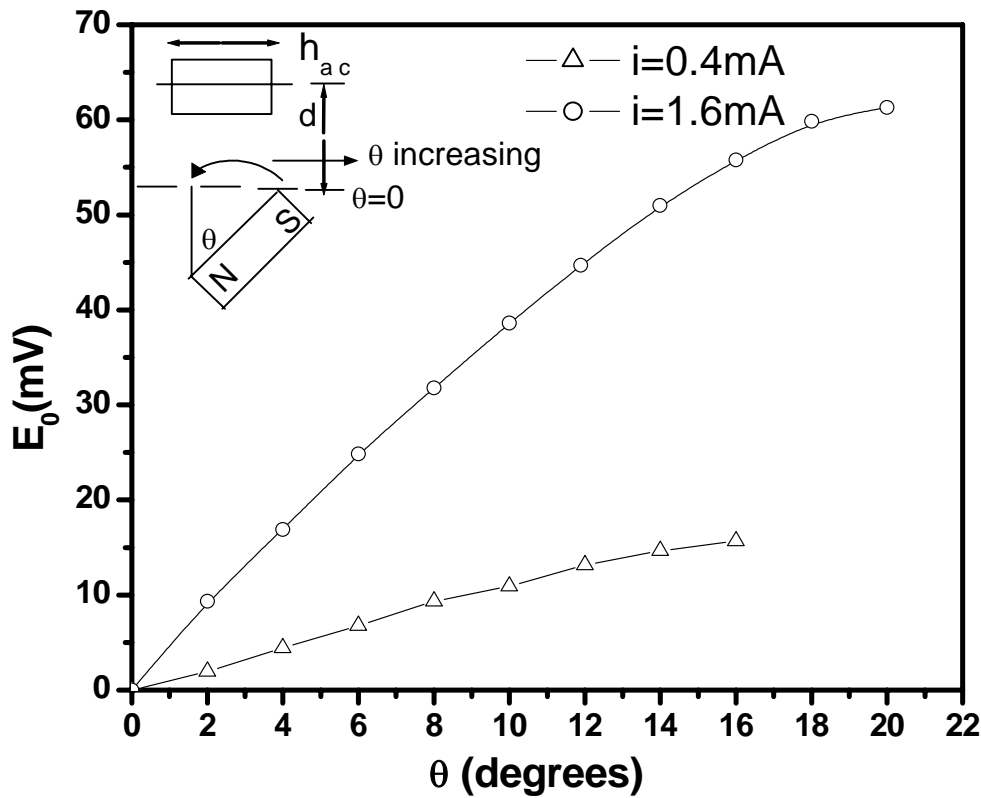

Fig (5)

Fig 5. The variation of output voltage $E_0$ with angular displacement θ of a bar magnet (d=5mm) in transverse configuration and at excitation currents of 0.4 and 1.6mA.

Two different values of excitation currents of 0.4mA and 1.6mA have been used as the parameters. The signal shows signs of saturation at large values of angular displacement of the magnet. In the intermediate positions, a linear region is obtained for both the values of excitation currents. We note that the magnitudes of the voltage responses are larger for higher values of excitation currents. The sensitivity of linear portions at $i_c$ = 0.4mA is 1.07mV/degree which increases to 3.4mV/degree at $i_c$ = 1.6mA. The range of sensitivity of the voltage response for $i_c$ = 0.4mA extends from $0<\theta<8^0$, which is less than that for $i_c$ = 1.6mA where it extends from $0<\theta<14^0$.

Fig.6 depicts the response of the transducer as an angular displacement sensor and in longitudinal configuration. Similar to the transverse configuration, here also the voltage response shows signs of saturation at higher values of angular displacements of the bar magnet.

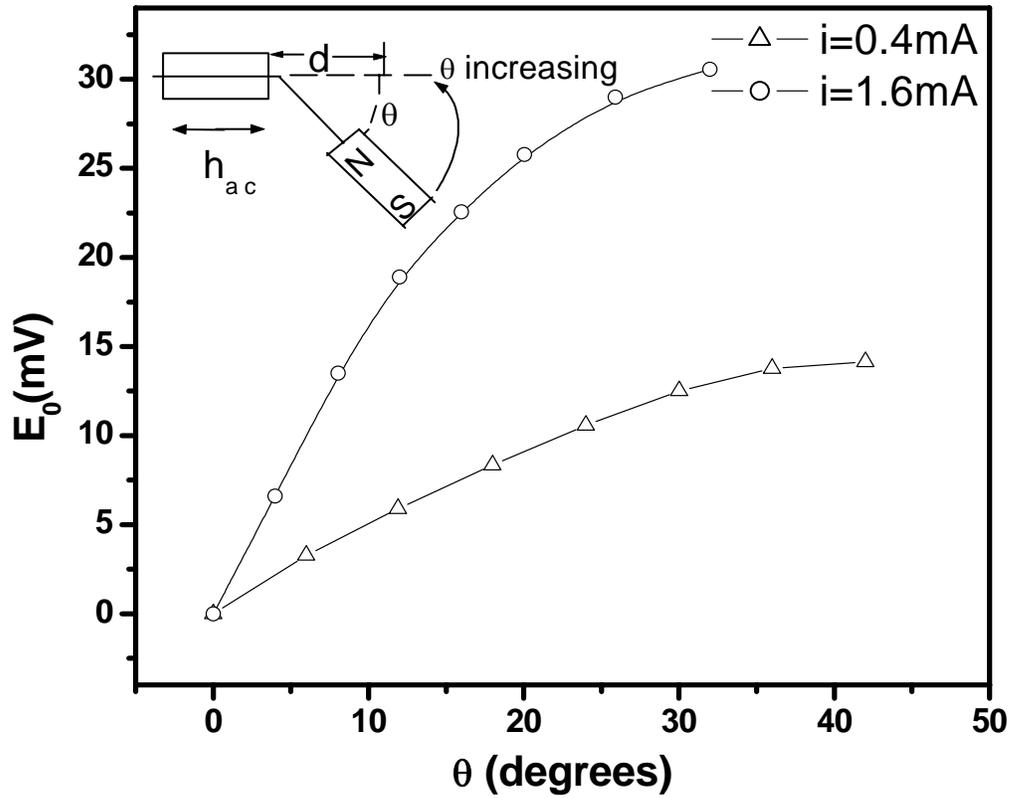

Fig (6)

Fig 6. The variation of output voltage $E_0$ with angular displacement θ of a bar magnet (d=5mm) in longitudinal configuration and at excitation currents of 0.4 and 1.6mA.

The sensitivity for $i_c$ = 0.4mA is about 0.38mV/degree which increases to 1.357mV/degree at higher value of excitation current, $i_c$ = 1.6mA. The range of linearity in this configuration corresponding to $i_c$ = 0.4mA is from $0<θ<8^0$ which is lower for higher value of excitation current, $i_c$ = 1.6mA, where it extends from $0<θ<11^0$.

Fig (7) depicts the hysteretic effects of the transducer as a linear displacement sensor in the longitudinal configuration and when the excitation current is fixed at 0.4mA.

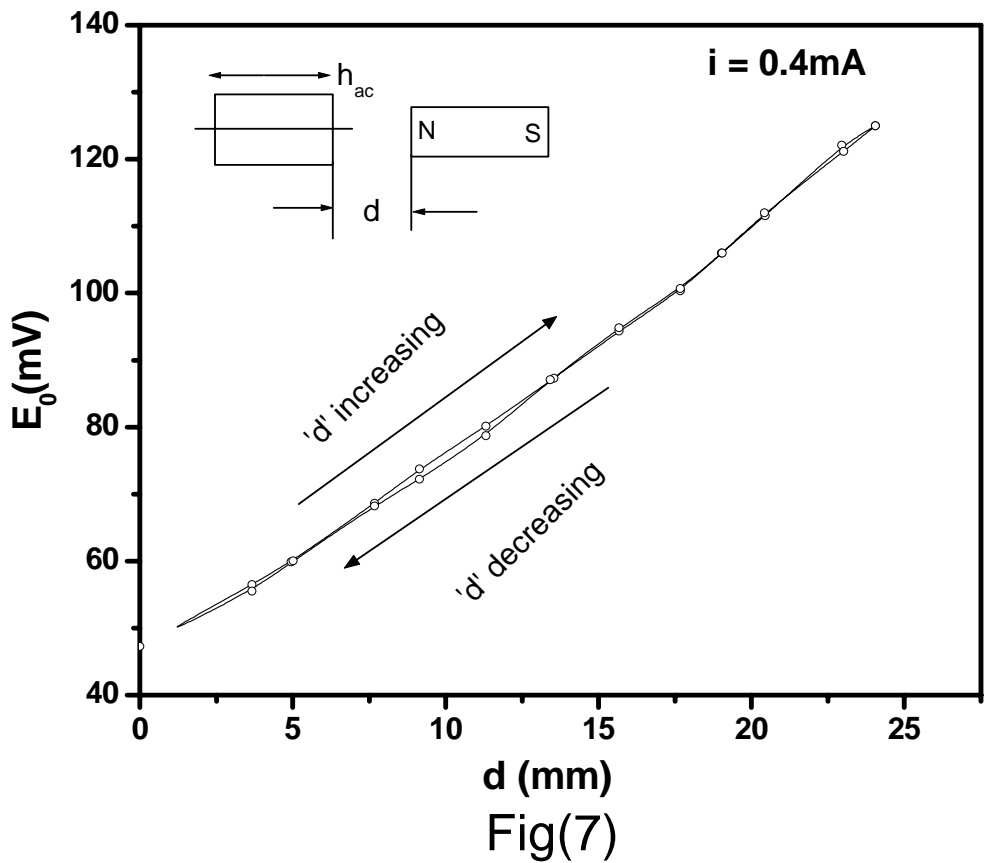

Fig(7)

Fig 7. Plot to show the hysteretic effects of the transducer as a linear displacement sensor in a longitudinal configuration. The voltage response was measured for increasing and decreasing values of the linear displacement 'd' of the bar magnet. The excitation current was maintained at 0.4mA.

It is clear from the figure that almost no changes in the sensitivity of the voltage response occur as the bar magnet is moved to and from the initial position.

## 4. Conclusions

A simple ac transducer assembly for field and displacement (linear and angular) sensor is presented. The transverse configuration is preferred over the longitudinal configuration owing to its high sensitivity in both the case of linear and angular displacement sensor. The results demonstrate the feasibility for using this system to detect small magnetic fields (less than 1Oe) and very small displacements of the order of a few micrometers.